\documentclass[letterpaper]{natureprintstyle}

\usepackage{amsfonts}
\usepackage{balance}
\usepackage{eurosym}

\usepackage{graphicx,epsfig}
\usepackage{amsmath,amssymb,bbm}
\usepackage{color}
\usepackage[usenames,dvipsnames]{xcolor}
\usepackage[colorlinks=true,linkcolor=Red,citecolor=Green,urlcolor=Black,linktoc=page]{hyperref}
\usepackage{multirow}
\usepackage[varg]{txfonts}

\usepackage[normalem]{ulem} 

\usepackage{fancyhdr}
\usepackage{mathtools}
\usepackage[caption=false]{subfig}
\usepackage{float}
\usepackage{flushend}

\DeclareCaptionLabelSeparator{csdel}{\,$\vert$\,}
\usepackage[figurename=Fig.,labelfont=bf,labelsep=csdel]{caption}

\usepackage[superscript,nomove]{cite}
 
\let\oldbibliography\thebibliography
\renewcommand{\thebibliography}[1]{%
  \oldbibliography{#1}%
  \setlength{\itemsep}{5pt}%
}
 
\usepackage[normalem]{ulem} 

\newcommand\red[1]{{#1}}  

\newcommand{\FG}[1]{{#1}}

\def\subinrm#1{\sb{\rm#1}}                                                       
{\catcode`\_=13 \global\let_=\subinrm}                                           
\mathcode`_="8000                                                                
\def\upsubscripts{\catcode`\_=12 }                                               

\upsubscripts

\begin{document}

\title{The ferroelectric field-effect transistor with negative capacitance}

\author
{I.\,Luk'yanchuk,$^{1,2}$ A.\,Razumnaya,$^{3,2}$ A.\,Sen\'{e},$^{1}$ Y.\,Tikhonov,$^{1,3}$ 
\& V.\,M.\,Vinokur$^{2\ast}$}

\date{
Integrating ferroelectric negative capacitance (NC) into the field-effect transistor (FET) promises to break fundamental limits of power dissipation known as Boltzmann tyranny. 
However, realizing the stable static negative capacitance in the non-transient non-hysteretic regime remains a daunting task.
The problem stems from the 
lack of understanding of how the fundamental origin of the NC due to the emergence of the domain state can be put in use for implementing the NC FET.
Here we put forth an ingenious design for the ferroelectric domain-based  field-effect transistor with the stable reversible static negative capacitance. Using dielectric coating of the ferroelectric capacitor enables the tunability of the negative capacitance improving tremendously the performance of the field-effect transistors.
}

\maketitle
\upsubscripts
\thispagestyle{fancy} 
\lfoot{\parbox{\textwidth}{ \vspace{0.3cm}
 \rule{\textwidth}{0.2pt}
\hspace{-0.2cm} \textsf{\scalefont{0.80}
    $^1$University of Picardie, Laboratory of Condensed Matter Physics, Amiens,
80039, France;
$^2$Terra Quantum AG, St. Gallerstrasse 16A, CH-9400 Rorschach, Switzerland.
    $^3$Faculty of Physics, Southern Federal University, 5 Zorge str., 344090 Rostov-on-Don, Russia;
    $^*$The correspondence should be sent to vv@terraquantum.swiss
} 
\vspace{-0.2cm}
\begin{center}{\scalefont{0.87} \thepage}\end{center}}} \cfoot{}

\vspace{-2cm}

\section*{INTRODUCTION}~~\newline
Dimensional scalability of field effect transistors (FETs) has reached the Boltzmann tyranny limit 
 because of transistors' inability to handle the generated heat\,\cite{Horowitz2005}. To reduce the power dissipation of electronics beyond this fundamental limits, negative 
capacitance (NC) of capacitors comprising ferroelectric materials has been proposed as a solution\,\cite{Salahuddin2008}.
The FET with a negative-capacitance ferroelectric layer has gained an enormous attention of researchers\,\cite{Tu2018,Kobayashi2018,Iniguez2019,Hoffmann2019,Alam2019,Bacharach2019,Cao2020,Hoffmann2021,Mikolajick2021,Khosla2021}. 
However, after impressive initial progress that has resulted in a rich lore massaging the aspects of technological benefits of the prospective stable static negative capacitance, the advancement in the field decelerated considerably. The lack of a clear self-consistent physical picture of the origin and mechanism of the stable static negative capacitance\,\cite{Alam2019,Hoffmann2020,Rethinking2020,Mikolajick2021,Khosla2021} not only retarded the craved technological progress, but has led to numerous invalid fabrications and misleading claims\,\cite{Cao2020}.

In this work, we put forth a foundational mechanism of the NC in ferroelectrics demonstrating inevitable emergence of the NC due to formation of polarization domains.  
We establish a practical design of the stable and reversible NC FET based on the domain layout. The proposed device is tunable and downscales to the 2.5-5\,nm technology node.

In what follows we review the state-of-the-art and basic concepts behind exploring ferroelectrics as the NC elements which constitute the base for our new results. We also mark the potential pitfalls in the NC implementing in the so far suggested NC FETs caused by depreciating the immanent role of domains.
Figure\,1 demonstrates the  principles of integrating the ferroelectric layer with the NC into the FET and the crucial role of domain states. The performance of the FET is quantified by the so-called subthreshold swing  
$S\!S=( {\partial \left( \mathrm{log}_{10}I_d\right) }/{\partial V_{g}})^{-1}$
 that describes the response of the drain current $I_d$ to the gate voltage $V_g$. The lower the value of the $S\!S$, the lower power the circuit consumes.  
In a  basic bulk metal-insulator-semiconductor field-effect transistor (MIS FET), shown in Fig.\,1a, which generalizes the MOSFET structure, 
the subthreshold swing is 
\begin{equation}
S\!S=\overset{S\!S_b}{\overbrace{\frac{\partial V_{s}}{\partial \left( \mathrm{log}_{10}I_d\right) }}}\overset{m}{\overbrace{\frac{\partial V_{g}}{\partial V_{s}}}},\quad m=1+\frac{C_{s}}{C_{g}}\,.
\label{MISFET}
\end{equation}
Here the first factor, $S\!S_b$, presents the response of  $I_d$ to the voltage $V_s$ at the conducting channel region, and the second factor, the so-called body factor, $m$, characterizes the response of the voltage $V_s$ to the applied voltage $V_g$.  Figure\,1a  shows the equivalent electronic circuit, with $C_g$\,and\,$C_s$ standing for the gate dielectric and semiconducting substrate capacitancies, respectively.  

The fundamental  constraint of the energy/power efficiency of the MIS FETs arises from  the thermal injection of electrons over an energy barrier enabling drain current flow and thus preventing the reduction of factor $S\!S_b$ below the $60$\,mV\,dec$^{-1}$, because the body factor $m>1$ at $C_g,C_s>0$.  
To overcome this limitation, the FETs incorporating the NC into its design has been proposed\,\cite{Salahuddin2008}. Indeed, replacing the gate dielectric with material with negative capacitance $C_{NC}$ would make the body factor $m<1$, thus, pushing the $S\!S$ below the Boltzmann limit.


\begin{figure*}[t!]
\center
\includegraphics [width=16cm] {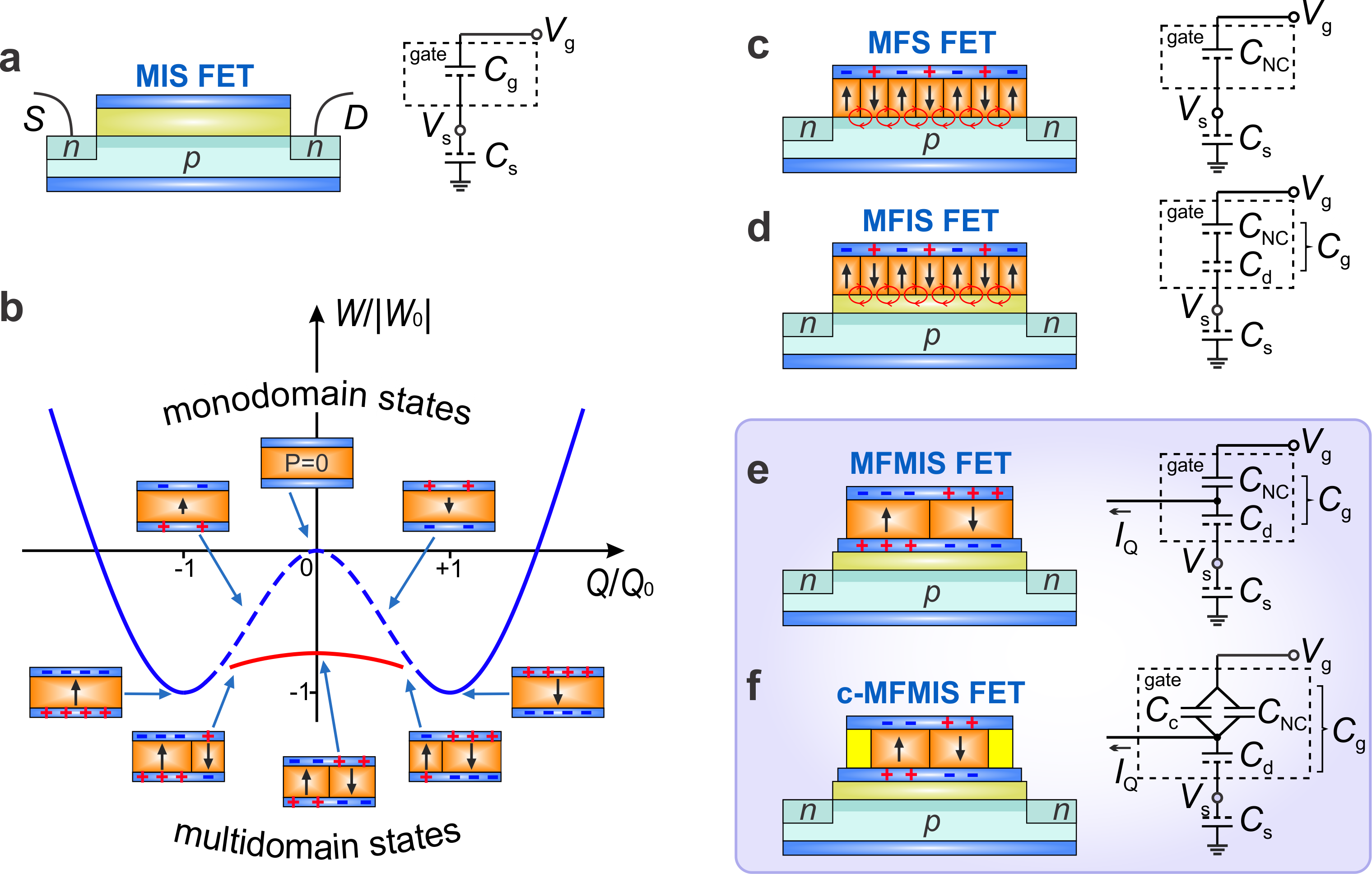}
\vspace{1cm}
\caption{ \textbf{Stability and feasibility of the ferroelectric-based NC FETs.}
\textbf{a,}\,Bulk MIS FET. Pale blue: p-type semiconducting substrate; dusty blue: source (S) and drain (D) of n-doped regions connected by conducting channel; sky blue: bottom ground electrode. The  gate stack: gate electrode and oxide gate dielectric layer (yellow green). Right: equivalent  circuit comprising gate dielectric capacitance, $C_g$, and capacitance to the ground, $C_s$, of the substrate. 
Solid lines depict electrodes, dashed lines depict interfaces without electrodes; $V_g$ is the applied gate voltage, $V_s$ is the channel potential.
 \textbf{b,}\,The normalized energy, $W/|W_0|$, 
 and polarization states of the ferroelectric (orange) capacitor as a function of the normalized driving charge $Q/Q_0$. Here $Q_0$ and $W_0$ are the equilibrium charge and energy of the monodomain short-circuited capacitor, respectively. 
 The unstable energy branch (dashed line) depicts the energy of the monodomain state. The energy of a stable two-domain state is shown by the red curve. Pluses and minuses show the distribution of charges at the plates. 
\textbf{c,}\,Multidomain structure of the MFS FET with redistributed electric charges at the top gate electrode and 
fringing electric fields (red loops) at the FS interface. Ferroelectric  capacitance $C_{NC}<0$ replaces $C_g>0$ of the MIS FET.
\textbf{d,}\, MFIS FET incorporates the dielectric layer with $C_d>0$. 
\textbf{e,}\, MFMIS FET integrating a floating gate electrode into the MFIS FET between ferroelectric and dielectric layers. 
\textbf{f,}\, The coated c-MFMIS FET: the dielectric shell (yellow) with  $C_c>0$ coats the ferroelectric layer.}
\label{FigFet}
\end{figure*}

Ferroelectric materials appear as best candidates for realizing negative capacitance in FETs\,\cite{Salahuddin2008}.  
The emergence of the NC in a ferroelectric capacitor follows from the Landau double-well landscape of the capacitor energy $W$ as a function of the applied charge $Q$\,\cite{Landauer1976}
(blue line in Fig.\,1b).  
The downward curvature of $W(Q)$ at small $Q$ 
 implies that the addition of a small charge to the ferroelectric capacitor plate, induces non-zero polarization and reduces its energy. Hence the negative value of the capacitance  $C_{NC}^{-1}=d^2W/dQ^2$.
Remarkably, even the domain formation due to 
 fundamental instability of a monodomain state\,\cite{Bratkovsky2000,Lukyanchuk2005,DeGuerville2005,Lukyanchuk2009,Zubko2010,2015Misirlioglu}, maintains the negative capacitance\,\cite{Bratkovsky2000,Zubko2016,Lukyanchuk2018,Lukyanchuk2019,Yadav2019,Das2021}. 
The energy $W(Q)$ of the multidomain state is lower then that of the monodomain state while the downward curvature at $Q=0$ is conserved, see the red line in\,Fig.\,1b illustrating an exemplary $W(Q)$ for the capacitor hosting two domains.
A detailed parsing of particularities related to the monodomain state instability and specific manifestations of the  multidomain state is presented in\,Supplementary Note\,1.

An inevitable multidomain formation posits the need for a detailed exploring possible ways of realization of the NC FETs.
While the multidomain configuration preserves the NC, the domain formation may trigger some undesired effects detrimental to realizing the NC FET. 
In particular, domains cause inhomogeneous charge\,\cite{2015Misirlioglu} and electric field\,\cite{Pavlenko2022} distribution, endangering the conducting channel in the most commonly discussed metal-ferroelectric-semiconducting MFS FETs (Fig.\,1c), as the voltage dispersion they cause becomes comparable with the transistor operation voltage. This problem can be mended by putting the ferroelectric layer into the pretransitional (incipient) regime just above the transition temperature, where the NC effect still persists, but the field-induced polarization  distribution is uniform\,\cite{Sene2010,Cano2010,Iniguez2019}.  However, this would limit the desirable 
decreasing of the body factor keeping it above 0.99\,\cite{Cano2010}. 
Another way out is introducing  an intermediate dielectric (insulating) buffer layer between the ferroelectric and semiconductor\,\cite{Hoffmann2021}, which corresponds to MFIS FET architecture shown in\,Fig.\,1d. This, in its turn, would 
not help much because smoothing the field nonuniformity would occur only at distances well exceeding the spatial scale  on which the  NC potential amplification effect is still actual.
A detailed analysis of particularities related to pitfalls of the discussed above architectures specific multidomain state is presented in\,Supplementary Note\,2.

The design  
with the floating gate electrode placed between ferroelectric and dielectric layers\,\cite{2014Frank,Khan2016} appeared to resolve this problem and to level the field inhomogeneities right below the electrode. The resulting  MFMIS structure is the conventional FET with the overimposed MFM capacitor, see\,Fig.\,1e. 
This architecture has attracted some critique\,\cite{Hoffmann2018nanoscale}, 
since it was believed that the anti-parallel domains formation inside the MFM capacitor would destabilize the NC. However, as we established in\,\cite{Lukyanchuk2019}, it is precisely the  two-domain configuration that provides the stable and operable NC 
 because of the possibility of manipulating the domain wall by the applied charges, not accounted for in\,\cite{Hoffmann2018nanoscale}. 

Here we introduce and devise the working regime of the MFMIS FET in which the NC effect  
emerges from the integrated MFM capacitor hosting two domains. We show that the MFMIS architecture not only free from the perils mentioned above 
but allows for an enormous improving MFMIS FET characteristics by coating the MFM capacitor with the dielectric capacitor in parallel connection.
The proposed coating design that we call c-MFMIS FET, shown in\,Fig.\,1f, provides degrees of freedom enabling a complete tunability of the dielectric parameters of the NC FET.


\begin{figure*}[t!]
\center
\includegraphics [width=16cm] {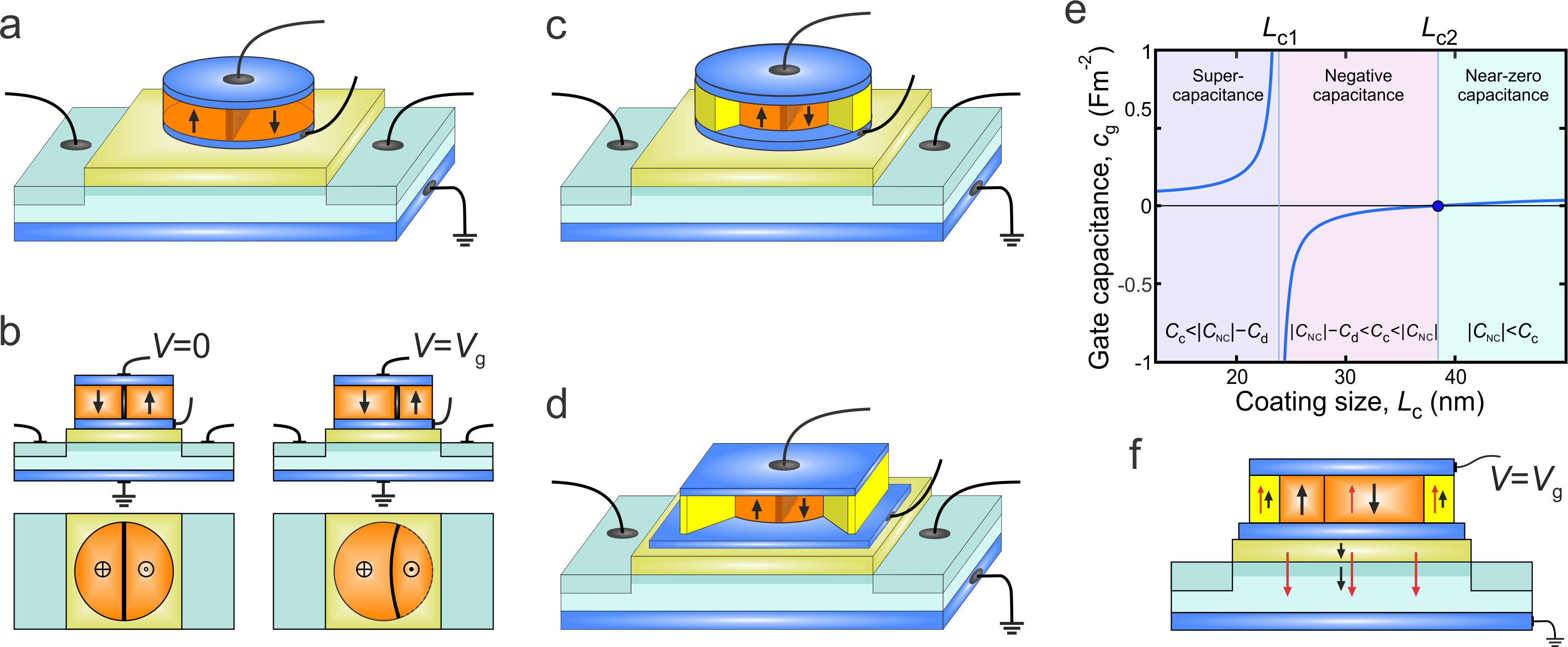}
\caption{ \textbf{Functioning of the MFMIS FET.} 
\textbf{a,}\,A three-dimensional sketch of the MFMIS FET comprising a disc-shape ferroelectric capacitor with the negative capacitance. 
\textbf{b,}\,Vertical and horizontal cross-sections of the MFMIS FET. At the zero gate voltage $V = 0$ (left hand side), the equilibrium location of the DW is in the
middle of the ferroelectric disc. At the finite applied voltage, $V = V_g$, the DW displaces from the middle and bends (right hand side).
\textbf{c,}\,A sketch of the coated MFMIS FET, in which the ferroelectric disc is sheathed by the dielectric shell. The dielectric shell shares the top-gate and floating-gate electrodes with the ferroelectric disc, hence making a complimentary capacitor going in parallel with the ferroelectric capacitor.   %
\textbf{d,}\,Rectangular modification of the c-MFMIS FET. The extended area of coating and gate dielectric layer capacitors  enhances the performance of the NC FET.
\textbf{e,}\,Normalized gate capacitance $c_g$ as function of the coating size $L_c$. Colored areas show three distinct regimes of the gate functioning.  
\textbf{f,}\,The sketch of the electric field (red arrows) and polarizations (black arrows) distribution inside the c-MFMIS FET upon applying voltage to the gate electrode.
}
\label{Fig3D}
\end{figure*}

\section*{RESULTS AND DISCUSSION}
\subsection{Two-domain negative capacitance of the MFM capacitor} ~~\newline

The nanodot-scale two-domain MFM is a major element of the MFMIS FET enabling the NC response via the charge-controlled motion of the DW. Following\,\cite{Lukyanchuk2019}, we discuss in detail the negative capacitance of the MFM capacitor, which is the base of our proposed device.
Shown in Fig.\,2a is the general view of MFMIS FET. Figure\,2b presents the vertical and horizontal cross-sections of a nanoscale ferroelectric disc-shape MFM capacitor integrated in the MFMIS FET.
At the zero charge $Q_f$ at the electrodes of the MFM capacitor, corresponding to the zero voltage at the transistor, the DW sits in the middle of the capacitor, see left hand side of Fig.\,2b. The intrinsic charges at the respective electrodes redistribute in order to compensate the depolarization charges of each domain (keeping the total charge $Q_f=0$) and to banish the electric field inside the ferroelectric disc, reducing thus the electrostatic energy. 
The finite charge $Q_f$, induced by the voltage $V=V_g$ applied to the transistor gate, displaces the DW from its middle zero-voltage position, see right hand side of  panels\,(b). Accordingly, the intrinsic charges rearrange (maintaining the total charge  $Q_f$) to compensate the depolarization fields of now unequal domains. 

The NC response arises since the length, hence the energy of the displacing DW, is sensitive to the shape of a ferroelectric capacitor. To ensure the best controlled performance of the NC, we choose a disc-like form of a capacitor. When
moving apart from the middle, the DW not only  compensates the electric field arising due to the charge transferred to the electrodes but, minimising its surface self-energy, shrinks in the width $w$ and bends because of the cylindrical shape of a ferroelectric. 
As a result, the DW overshoots towards the edge beyond the electrostatics-demanded equilibrium position at which the internal electric field would have disappeared. Hence the net electric field does not vanish but flips over and goes from the negatively charged electrode to the positively charged one. This counterintuitive outcome precisely expresses the phenomenon of the negative capacitance. 
Note, however, that at some threshold value of the applied charge, $Q_f^*$, when it becomes approximately equal to the depolarization charge of the uniformly oriented polarization, $Q_0$, the DW reaches the edge of the ferroelectric layer and leaves the sample.  The monodomain state with positive capacitance restores; this corresponds to the termination of the red branch in Fig.\,1b. 

The quantitative description of the NC of the disk-shape two-domain ferroelectric capacitor is given by\,\cite{Lukyanchuk2019}
\begin{equation}
C_{NC}=-\gamma_2 \frac{D_{f}}{\xi_{0}}C_f, 
\label{CNC}
\end{equation} 
where we explicitly spotlighted the capacitance $C_f=\varepsilon_{0}\varepsilon_{f} S_f /d_f>0$, which is the capacitance of the monodomain MFM capacitor in the stable state at 
$Q=\pm Q_0=\pm S_f P_0$ (minima of the $W(Q)$ dependence in Fig.\,1b). The negative factor, $-\gamma_2 D_f/\xi_0$,
reflects the features brought in by the DW displacement in the two-domain configuration.
Here $D_f$ and $d_f$ are the diameter and height of the capacitor, respectively,  $S_f=\pi D_f^2/4$ is the area of the ferroelectric  plate surfaces, and $P_0$ is the polarization of the ferroelectric in the equilibrium state. The coherence length $\xi_0\simeq 1$nm describes the DW thickness, the dimensionless geometric factor $\gamma_2\approx 4.24$ reflects the internal profile of polarization inside the DW and the DW bending in the cylindrical gate, and $\varepsilon_{f}$ and $\varepsilon_{0}$ are the dielectric constant of the ferroelectric material and the vacuum permittivity, respectively.

Figure\,1b displays the energy advantage of the two-domain state (whose energy is shown by the red curve), with respect to the  usually considered uniform NC state  (the blue dashed curve), both states preserving the same charge $Q$ at the electrodes. 
To create the two-domain state from the uniformly-polarized state one has to suppress polarization in a fraction of the ferroelectric occupied by the DW, while to depolarize the monodomain state by the electric field due to the uniformly distributed  charge $Q$, one would have had to suppress the ferroelectricity within the whole volume which is much more energetically costly.

As a next step, we integrate the MFM two-domain NC-capacitor into the MFMIS FET architecture.  

\bigskip
\subsection{The MFMIS FET}~~\newline

This device comprises the gate stack overimposed on a semiconducting substrate  in which the source and drain parts are connected by the gate-operated conducting channel, see\,Fig.\,2a. The gate stack includes the MFM capacitor and the gate 
insulating layer separating it from the substrate. This is the high-$\kappa$ dielectric layer, preventing a charge leakage between the lower capacitor's plate and the semiconducting channel.


\begin{figure*}[t!]
\center
\includegraphics [width=18cm] {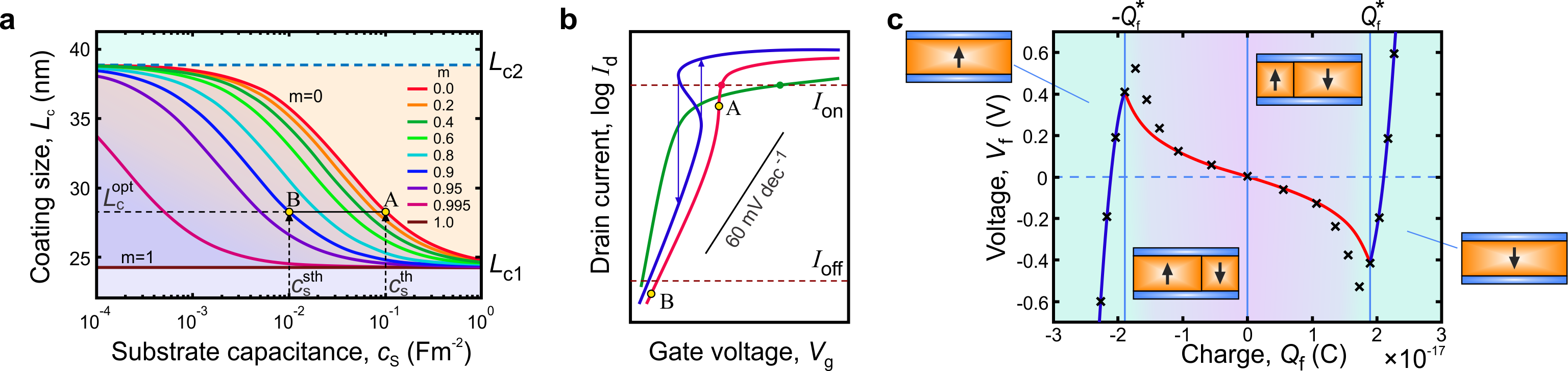}
\caption{\small  \textbf{Practical design of the c-MFMIS FET.} 
\textbf{a,}\,A set of $L_c(c_s)$ dependencies for determining the optimal parameters for the gate-substrate match in the c-MFMIS FET.  
The arrows define the procedure for selecting the optimal coating size $L_c^{\mathrm{opt}}$ providing the best performance and stability of the transistor. 
\textbf{b,}\,Transport characteristics $I_d(V_g)$ of the representative c-MFMIS FETs. The red curve shows $I_d(V_g)$ for the optimally designed c-MFMIS FET. 
The green and blue curves show the $I_d(V_g)$ dependencies for transistors with the steeper $S\!S$ having the voltage-independent (green) and nonlinear in voltage (blue, with hysteresis) substrate capacitancies. 
\textbf{c,}\,The charge-voltage characteristics, $V_f(Q_f)$, of the disc-shape MFM capacitor. The monodomain and two-domain operational regions are displayed as the turquoise and purple, respectively, inter-shaded regions. The threshold charges $\pm Q_f^*$ at which the domain wall leaves the ferroelectric nanodot, mark the transition from the negative capacitance two-domain region (red curve) to the positive capacitance monodomain region (blue curves). The crosses stand for the results of the numerical simulations.}
\label{FigMatch}
\end{figure*}

The top MFM capacitor plate is the gate electrode connecting the transistor to the external voltage source. The bottom capacitor electrode is an intermediate electrically isolated floating gate electrode of the transistor that preserves the entire charge, most commonly the zero total charge, constant, stabilizing the ferroelectric two-domain state.  
Furthermore,
the floating gate makes the potential along the ferroelectric interface even, maintaining, therefore, a uniform electric field across the gate stack and substrate. 
Along the way, the floating gate resolves a frequent issue of neutralizing the parasitic charges that may be trapped by interfaces during the fabrication and functioning.  
Maintaining the working charge and providing a regular rubbing out the parasite leaking charges
with the removal time faster than the leakage time\,\cite{Khan2016} 
is achieved via the standard discharging methods. For instance, it is implemented by either harnessing
the Fowler–Nordheim tunneling and the hot electrons injection\,\cite{Hasler2001}, or by
circuiting the gate by the auxiliary charge-carrying current, $I_Q$, contact, see Fig.\,1e, to a certain source for a given combination of electric inputs ensuring the proper sequence of discharging and high resistance modes\,\cite{Kotani1998}.

An effective electronic circuit of the MFMIS FET, shown in Fig.\,1e, is similar to that of the multidomain MFIS FET (Fig.\,1d). The difference is that now the task of leveling the depolarization field inhomogeneities is taken by the floating gate electrode. Therefore, the gate dielectric layer can be safely engineered as an utmostly thin one, down to the technologically-acceptable limit of a few nanometers. This increases $C_d$ with respect to $|C_{NC}|$, opening doors to making the gate capacitance negative, $C_g^{-1}=C_{d}^{-1}-|C_{NC}^{-1}|<0$. Yet it is hard to achieve a required largeness of $C_d$ with respect to $|C_{NC}|$ due to the 
restrictions imposed by materials compatible with the silicon CMOS technologies. To meet the challenge, we devise a coated c-MFMIS FET that critically changes the situation and breaks ground for the unlimited enhancement of the performance of the NC FET transistor. 

\bigskip
\subsection{The c-MFMIS FET}~~\newline
The coating of the ferroelectric layer with the dielectric oxide sheath confined by the same electrodes, see Fig.\,2c,d, straightforwardly incorporates an additional capacitor with the capacitance $C_c>0$ in parallel to $C_{NC}$. This results in a radical improvement of the controlled tuning of the gate capacitance. In particular, manipulating with the sizes of the coating oxide layer and with the geometrical design of the device as a whole, provides a broad variation in its performance characteristics and functioning regimes. The panels (c) and (d) exemplify possible designs. The panel (c) shows an annulus-like coating capacitor, while panel (d) displays a rectangular design of the coating layer and, in addition, the possibility of increasing the area of the floating gate electrode with respect to the top gate electrode. The latter design allows for controllable increasing capacitances of the coating and gate-dielectric layers most efficiently, maintaining the miniaturization of the device. Note, that in all the geometries, the core ferroelectric should maintain its disc-like shape ensuring the optimal manipulation with the DW.

Shown in Fig.\,1f is the equivalent circuit of c-MFIS FET. The important advance is that the gate capacitance becomes 
\begin{equation}
C_g=\frac{1}{C_d^{-1}+(C_c-|C_{NC}|)^{-1}}, 
\label{cGate}
\end{equation}
which permits tuning $C_g$ over the widest range of values by the appropriate modifying the parameters of the coating capacitor. 
We reveal the rich
dependence of $C_g$ on the coating layer size, $L_c$ choosing the rectangular geometry of the coating layer (Fig.\,2c). 
The capacitance of the disk-shape two-domain ferroelectric capacitor, $C_{NC}$, is given by Equation\,(\ref{CNC}). 
The capacitances of the coating and gate dielectric layers are taken in a standard form as
$C_{c}=\varepsilon _{0}\varepsilon _{c} S_c /d_c$ and $C_{d}=\varepsilon _{0}\varepsilon _{d} S_d /d_d$, respectively. Here $S_f=\pi D_f^2/4$, $S_c=  L_c^2 - S_f$ and $S_d=L_d^2$ are areas of the ferroelectric, coating dielectric and gate dielectric plate surfaces, respectively. 

The behavior of $C_g$ defined by Equation\,(\ref{cGate}) is most generic and does not depend critically on the specific choice of materials. 
For practical applications, we choose both, the coating layer and the gate-dielectric layer, be composed of the Si-compatible dielectric  HfO$_2$ with the respective dielectric constants $\varepsilon_c$, $\varepsilon_d$ being approximately equal to 25. 
The ferroelectric disc of the diameter $D_f\simeq 6$\,nm and thickness $d_f\simeq 3$\,nm, can be fabricated out of the ferroelectric phase of HfO$_2$ or its Zr-based modification, Hf$_{0.5}$Zr$_{0.5}$O$_2$ with $\varepsilon_f\simeq 50$\,\cite{Hoffmann2020}. 
In fact, $\varepsilon_f$ is the only relevant material parameter that defines the NC properties of the two-domain ferrolectric layer; therefore, the consideration applies equally well to other similar ferroelectrics, for instance, to perovskite oxides, 
like strained PbTiO$_3$ with about the same $\varepsilon_f$. The height  of the gate dielectric layer is taken as $d_d\simeq 3$\,nm, whereas the thickness of the coating layer, $d_c$, is equal to $d_f$. The size of the rectangular floating-gate electrode, $L_d$, defining the size of the gate dielectric capacitor, is taken as $1.2L_c$. 

Figure\,2e displays the derived normalized gate capacitance, $c_g=C_g/S_d$ as function of $L_c$. The presented $c_g(L_c)$ dependencies are the  Equation\,(\ref{cGate}) plots, into which the given above capacitancies, $C_{NC}$, $C_{c}(L_c)$ and $C_{d}(L_c)$ are substituted.  Looking at the plots, one discriminates 
the three distinct regimes of the gate functioning set by two critical sizes, $L_{c1}<L_{c2}$, of the coating layer: (i) the super-capacitance, $L_{c}<L_{c1}$, (ii) the negative-capacitance, $L_{c1}<L_{c}<L_{c2}$, and (iii) the near-zero capacitance, $L_{c2}<L_{c}$. All these three  distinct regimes are of tremendous relevance for applications. Below, we restrict ourselves to the detailed analysis of the NC regime,  i.e., the regime where $c_g<0$, leaving the detailed discussion of regimes (i) and (iii) to forthcoming publication.
Note that although in the NC regime the average polarization of the ferroelectric nanodot is aligned with the voltage drop across the capacitor, the electric field inside the coated layer is directed oppositely to it. Accordingly, the polarisation induced inside the dielectric oxide sheath is opposite to the polarization of the ferroelectric nanodot, see Fig.\,2f. It is important that the absolute value of the NC capacitance can be done arbitrary small on approach to the resonance regime $|C_{NC}|=C_c$ of Equation\,(3), i.e., upon $L - L_{c2}\to 0^{-}$. In this regime the nonlinear effects in the $Q-V$ characteristics of the MFM capacitor become of prime relevance.

An unrestricted range of variation of $c_g$, from minus infinity to  zero, as seen from Fig.\,2e, allows for the unlimited tuning of the magnitude of the gate NC. In particular, the possibility of making it arbitrary small, enables us to overcome a previously insurmountable obstacle of the proper matching between the gate, $C_g$, and substrate, $C_s$, capacitancies and obtain the desirable value of the body factor
\begin{equation}
m=1+\frac{C_s}{C_g}=1+\frac{C_s}{C_d} +\frac{C_s}{C_c-|C_{NC}|}\,, 
\label{bodym}
\end{equation} 
within the NC FET operational interval $0<m<1$, 
getting thus the remarkably low values of  $S\!S$\,(\ref{MISFET}), the task not achievable by previously suggested architectures.

Now the task is to find the optimal coating size $L_c$, given the normalized substrate capacitance $c_s=C_s/S_d$, that ensures matching to targeted value of $m$.
To that end, we employ Equation\,(\ref{bodym}) which defines the implicit $L_c$ dependence of $c_s$ at given $m$. 
The shown in Fig.\,3a family of $L_c (c_s)$ curves for different $m$, confined between $m=0$ (red) and $m=1$ (brown) characteristics, represents the stable working interval for our exemplary c-MFMIS FET.

The region below the brown line where $m$$>$$1$, i.e., $S\!S>60$\,mV\,dec$^{-1}$, corresponds
to small sizes of the coating layer, $L_c<L_{c1}$. 
The region above the $m=0$ curve but below the $L_c=L_{c2}$ line is the hysteretic loss of the reversibility region. 
Therefore, the proper choice of $L_c$, in the interval $L_{c1}<L_c<L_{c2}$, enables the desired magnitude of $m$ for a given value of the substrate capacitance $c_s$. Although in our model example, the lateral size of the coating layer varies between $L_{c1}\approx 24$\,nm and $L_{c2}\approx 38$\,nm,
it can be significantly reduced down to practically the diameter of the ferroelectric disc, by increasing the dielectric constant of the coating material by factor of four.

\bigskip
\subsection{Practical design. Optimal match between gate and substrate}~~\newline
The established characteristics of the c-MFMIS FET enable us to go beyond the past \textit{prima facie} technological concepts and turn to the practical design in relevant industrial environment. 
The critical challenge\,\cite{Cao2020} emerging when engineering the NC FET, is the complex and highly nonlinear nature of $C_s$ since the latter comprises contributions from the capacitance of the depleted layer, from the quantum capacitance due to charge carriers injected into the conducting channel, from the interface charges, and, finally, from the source/drain geometrical capacitancies.  While being of a relatively low value when coming mainly from the capacitance of the depletion layer in the low-conducting regime at small voltages, $C_s$ increases dramatically, typically by order of magnitude, due to the injection of conducting electrons into the channel caused by the gate bias near and above the threshold value. 

Using our established concept of $L_c(c_s,m)$ characteristics, Fig.\,3a,
enables the determining an optimal match between the NC gate capacitance and the semiconducting substrate capacitance ensuring the best-performance $S\!S$ and stability of the transistor.
To exemplify the nonlinear behaviour of the substrate, we take the normalized capacitance $c_s^{\mathrm{sth}}\simeq 10^{-2}$\,Fm$^{-2}$ in the low-voltage subthreshold regime and $c_s^{\mathrm{th}}\simeq 10^{-1}$\,Fm$^{-2}$ in the high-voltage near-threshold regime. 
Next, we set the condition $m=0$ at the steepest point of the $I_d(V_g)$ curve, i.e., in the near-threshold gate voltage where $c_s=c_s^{\mathrm{th}}$. This condition 
visualized by the point A in Fig.\,3a, provides us with the optimal value of the size of the coating layer, $L_c^{\mathrm{opt}}\approx 28$\,nm. 
Decreasing the gate voltage $V_g$ to subthreshold values, reduces $c_s$ and moves it to the  left from the point A along the black line until reaching the point B 
corresponding to  $c_s=c_s^{\mathrm{sth}}$ at $V_g\approx 0$. 
The body factor at the point B is $m=0.9$ which gives  {$S\!S\approx 54$\,mV\,dec$^{-1}$}. 

The possible transfer $I_d(V_g)$ characteristics\,\cite{Cao2020} are schematically illustrated in Fig.\,3b. The optimal characteristic (red line) derived according to the devised above operating procedure starts with the relatively modest $S\!S\approx 54$\,mV\,dec$^{-1}$, steepens upon the increase in $V_g$, and reaches its steepest value in the near-threshold region. 
The location of points A and B corresponds to the capacitancies $c_s^\mathrm{{th}}$ and $c_s^\mathrm{{sth}}$ of panel\,(a). 
The optimal regime maintains the steep slope of the $I_d(V_g)$ dependence over the entire voltage working range and includes not only the subthreshold, but also near- and above threshold regimes, preserving the stable and hysteresis free $I$-$V_g$ transfer characteristics at the same time.

Because of the nonlinearity of $c_s$, the transfer characteristics of the designed c-MFMIS FET with $L_c=L_c^{\mathrm{opt}}$ demonstrates the better performance at the same on-off current switching ratio, $I_{on}/I_{off}$, than the shown by the green curve commonly assumed NC FET with the voltage-independent substrate capacitance $c_s$ (although having the steeper initial $S\!S$). 
At the same time, an attempt to engineer the NC FET having the   
nonlinear $c_s$ with an initially steeper $S\!S$ (exemplified by the blue curve), results in the hysteretic switching instability.  

\smallskip

\subsection{Practical design. Compact gate model}~~\newline
In order to provide incorporating the two-domain c-MFMIS FET architecture into the industrially-standardized circuiting, we design the  scalable compact model describing the $Q_{g}$-$V_{g}$ characteristics of the coated NC gate stack.
In the low-voltage and low-charge
operational mode of the NC FET, this compact model is defined by the linear relation $Q_{g}=C_{g}V_{g}$, where $C_{g}$ is given by Eqs.\,(2) and (3). Looking forward to extensive applications of our compact model for the description of the c-MFMIS FET, we expand the compact model's working range over to the nonlinear regime
where substantial shifts of the domain wall and even its escape from the sample may occur.

The complete set of the $Q_{g}$-$V_{g}$ characteristics of the gate is defined by the individual electric properties of its components, including the gate dielectric capacitor, coating capacitor, and the
MFM capacitor which form the equivalent circuit shown in Fig.\,1f and which are
characterized by the linear, $V_{d}=C_{d}^{-1}Q_{d}$, $V_{c}=C_{c}^{-1}Q_{c}$, and nonlinear, $V_{f}=V_{f}\left( Q_{f}\right) $, constitutive relations
respectively. 
In general, the $V_{g}$-$Q_{g}$ characteristics can be
parametrically plotted as functions of the running parameter  $Q_{f}$
\begin{gather}
V_{g}=\left( 1+\frac{C_{c}}{C_{d}}\right) V_{f}(Q_{f})+\frac{Q_{f}}{C_{d}}
\label{param} \\
Q_{g}=C_{c}V_{f}(Q_{f})+Q_{f}\,,  \notag
\end{gather}
based on the relations $Q_{g}=Q_{d}=Q_{c}+Q_{f}$, $V_{g}=V_{d}+V_{f}
$, and $V_{f}=V_{d}$ for the circuit  in\,Fig.\,1f.

\renewcommand{\arraystretch}{2}
\begin{table*}[h!]
\centering
\begin{tabular}{|c|c|c|c|}
\hline\hline
Type of \ capacitor & Capacitance, NC & Energy at $Q=0$  & Reference \\ 
\hline\hline
\multicolumn{1}{|l|}{MFM monodomain (unstable)} & \multicolumn{1}{|l|}{$%
C_{1}=-\gamma _{1}C_{f}\simeq -2\,C_{f}\,$} & \multicolumn{1}{|l|}{$W_{1}=0
$} &  \\ \hline
\multicolumn{1}{|l|}{MFM two-domain (stable)} & \multicolumn{1}{|l|}{$%
C_{2}=-\gamma _{2}\frac{D}{\xi _{0}}C_{f}\,\simeq -25.4\,C_{f}\,$} & 
\multicolumn{1}{|l|}{$W_{2}/\left\vert W_{0}\right\vert =-1+\frac{64\nu }{%
\pi }\frac{\xi _{0}}{D}\simeq -0.66$} & \cite{Lukyanchuk2019} \\ \hline
\multicolumn{1}{|l|}{MFSM multidomain (stable)} & \multicolumn{1}{|l|}{$%
C_{m}=-\gamma _{m}\frac{d_{f}}{w}C_{f}\simeq -4.55\,C_{f}$} & 
\multicolumn{1}{|l|}{$W_{m}/\left\vert W_{0}\right\vert =-1+\frac{7\zeta (3)%
}{\pi ^{3}}\frac{\kappa }{\varsigma }\frac{w}{d_{f}}\simeq -0.46$} & \cite{Lukyanchuk2018} \\ 
\hline\hline
\end{tabular}
\caption*{\textbf{Supplementary Table I}}
\end{table*}

The nonlinear constitutive relation, $V_{f}\left( Q_{f}\right)$, of the MFM  capacitor is the core relation that defines different regimes of the gate functioning. 
Shown in the\,Fig.\,3c, are the results of the phase-field simulations (crosses), see Methods, and the analytical outcome of the developed scalable compact model (solid lines).  The $Q_{f}$-$V_{f}$ characteristic reveals two different operational modes of the MFM capacitors. The NC low-charge branch, $V_{2}\left( Q_{f}\right)$ (red curve), corresponds to the two-domain state where the domain wall motion is responsible for the electric properties of capacitor. The high-charge branch,  $V_{1}\left( Q_{f}\right)$  (blue curve), with the positive differential capacitance, $C_f=dV/dQ>0$, corresponds to the monodomain state  where the domain wall is gone.   
As a result, the $Q_{f}$-$V_{f}$ characteristic of the ferroelectric capacitor is presented by the synthetic dependence 
\begin{equation}
V_{f}\left( Q_{f}\right) = \begin{cases}
V_{2}(Q_{f}) & \text{if $ \left\vert
Q_{f}\right\vert <Q_f^{\ast }$} \\
V_1(Q_{f}) & \text{if $ \left\vert
Q_{f}\right\vert >Q_f^{\ast }$}
\end{cases}
\label{cases}
\end{equation}
where $Q_f^{\ast }$ is the charge at which the branches $V_{2}\left( Q_{f}\right)$ and $V_{1}\left( Q_{f}\right)$ meet. 

The corresponding to the monodomain state branch of the $V_{f}$-$Q_f$ characteristic, is given by the parametric dependence of $V_1(Q_f)$ upon the polarization $P$,
\begin{gather}
V_1(P)=-d_{f}\left( 2a^*_{3}P+4a^*_{33}P^{3}+6a_{333}P^{5}\right) \notag \\
Q(P)=-S_{f}\left(
P-\varepsilon _{0}\varepsilon _{i}\frac{V_1(P)}{d_{f}}\right)\,,
\label{VP}
\end{gather}
derived from the uniform Ginzburg-Landau equation.
The coefficients $a^*_{3}$, $a^*_{33}$ and $a_{333}$, and the background dielectric constant $\varepsilon_i$ are defended in Methods.  

For the $V_{2}(Q_f)$ function describing the two-domain case we use the analytical approximation\,\cite{Lukyanchuk2019}
\begin{equation}
V_{2}(Q_{f})\approx-\frac{0.27}{\psi \left( Q_{f}/Q_{0}\right)}\frac{Q_f}{C_f}\frac{\xi
_{0}}{D},  
\label{Vf}
\end{equation}%
where $\psi(s)$ ($0<s<1$), introduced in\,\cite{Lukyanchuk2019}, is the function accounting for the geometry of the system which we fit by
\begin{equation}
\psi (s)\approx \left( 1.0-0.027s-0.95s^{2}-0.34s^{3}+0.32s^{4}\right) ^{1/2}. 
\label{approxs}
\end{equation}
In the linear in $Q_f$ approximation, where $s\rightarrow 0$, Equation\,(\ref{Vf}) gives $C_{NC}$ in Equation\,(\ref{CNC}). 

Combining branches given by Eqs.\,(\ref{VP}) and (\ref{Vf}) provides an excellent approximation for the results of the numerical simulations of the compact model, see Fig.\,3c. The slight overshoots at $\pm Q_f^*$ correspond to numerical singularities appearing at the moments where the DW leaves the MFM capacitor. 

To summarize, the achieved understanding that the fundamental mechanisms of the NC is the domain action, bestows closing the gap between the concept of the ferroelectric negative capacitance and its realization in electronic devices. 
It enables designing a stable NC-based FET, whose coating-shell architecture of the gate promises a notable enhancement of prospective performance 
and high tunability of characteristics allowing the perfect match with advanced FET architectures. Our findings lay out the way for scaling the NC FET nanoelectronics down to 2.5-5\,nm technology nodes via utilizing the CMOS-compatible ultra-thin and ultra-small ferroelectric disc as a core of the NC gate.

\section*{METHODS}~~\newline
\subsection{Functional}~~\newline
\medskip
\small{ 
To carry out the numerical modelling of polarization structures in a ferroelectric layer, we use the  most thoroughly studied free energy functional for the PbTiO$_3$, 
\begin{equation}
\begin{aligned} 
   F =  \int \biggl( \Bigl[ a_{\mathit{i}}^{*}(u_{m}, T)P_{\mathit{i}}^{2} &+ a_{\mathit{ij}}^{*}P_{\mathit{i}}^{2}P_{\mathit{j}}^{2}
   + a_{\mathit{ijk}}P_{\mathit{i}}^{2}P_{\mathit{j}}^{2}P_{\mathit{k}}^{2} \Bigr]_{\mathit{i \leq j \leq k}}  \\
   &+ \frac{1}{2}G_{\mathit{ijkl}}(\partial_{\mathit{i}}P_{\mathit{j}})(\partial_{\mathit{k}}P_{\mathit{l}})
   + (\partial_{\mathit{i}}\varphi)P_{\mathit{i}} - \frac{1}{2} \varepsilon_{0} \varepsilon_{i} (\nabla \varphi)^{2} \biggr)\,d^{3}r, 
\end{aligned}
\label{FEFunctional}
\end{equation}
where the sum is taken over the cyclically permutated indices $\{i,j,k,l\}=\{1,2,3\}$ (or $\{x,y,z\}$).
Functional\,(\ref{FEFunctional})  includes the Ginzburg-Landau (GL) energy of the strained ferroelectric layer\,\cite{Pertsev1998} written in a form given in\,\cite{Baudry2017} (the square-bracketed term), the polarization gradient energy\,\cite{Li2002} (the term with coefficients $G_{\mathit{ijkl}}$), and the electrostatic energy, including the coupling of polarization with electric field\,\cite{Lukyanchuk2009}, $E_{\mathit{i}}=-\partial_{\mathit{i}} \varphi$, described through the electrostatic potential $\varphi$ (the two last terms). 
The strain-renormalized GL coefficients for the PbTiO$_3$ layer (accounting partially for the elastic energy) are taken as\,\cite{Pertsev1998}, 
$a_{1}^{*}$,$a_{2}^{*}$\,=\,$3.8\times10^5(T-479 ^{\circ}C)-11\times10^{9}u_{m}$\,C$^{-2}$m$^{2}$N$^{-1}$,
$a_{3}^{*}$\,=\,$3.8\times10^5(T-479^{\circ}C)+9.5\times10^{9}u_{m}$\,C$^{-2}$m$^{2}$N$^{-1}$,
$a_{11}^{*}$, $a_{22}^{*}$\,=\,$0.42\times10^{9}$\,C$^{-4}$m$^{6}$N,
$a_{33}^{*}$\,=\,$0.05\times10^{9}$\,C$^{-4}$m$^{6}$N,
$a_{13}^{*}$,$a_{23}^{*}$\,=\,$0.45\times10^{9}$\,C$^{-4}$m$^{6}$N,
$a_{12}^{*}$\,=\,$0.73\times10^{9}$\,C$^{-4}$m$^{6}$N,
$a_{111}$, $a_{222}$, $a_{333}$\,=\,$0.26\times10^{9}$\,C$^{-6}$m$^{10}$N,
and $a_{123}$\,=\,$-3.7\times10^{9}$\,C$^{-6}$m$^{10}$N. The 
misfit strain is taken as  $u_{m}=-0.013$.
The gradient coefficients are taken as for PbTiO$_3$ bulk material\,\cite{Li2002} (with all possible cubic index permutations),   
$G_{1111}$\,=\,$2.77\times10^{-10}$\,C$^{-2}$m$^{4}$N,
$G_{1122}$\,=\,0.0,
and $G_{1212}$\,=\,$1.38\times10^{-10}$\,C$^{-2}$m$^{4}$N.
The background dielectric constant of the non-polar ions was taken as $\varepsilon_i \simeq 10$\,\cite{Mokry2016} for PbTiO$_3$ and $\varepsilon_i \simeq 25$ for semiconducting layer. The vacuum permittivity is $\varepsilon_{0}=8.85\times 10^{-12}$\,CV$^{-1}$m$^{-1}$.
  
}

\subsection{Phase-field simulations}
\begin{figure}[t!]
    \centering
    \includegraphics[width=0.475\textwidth]{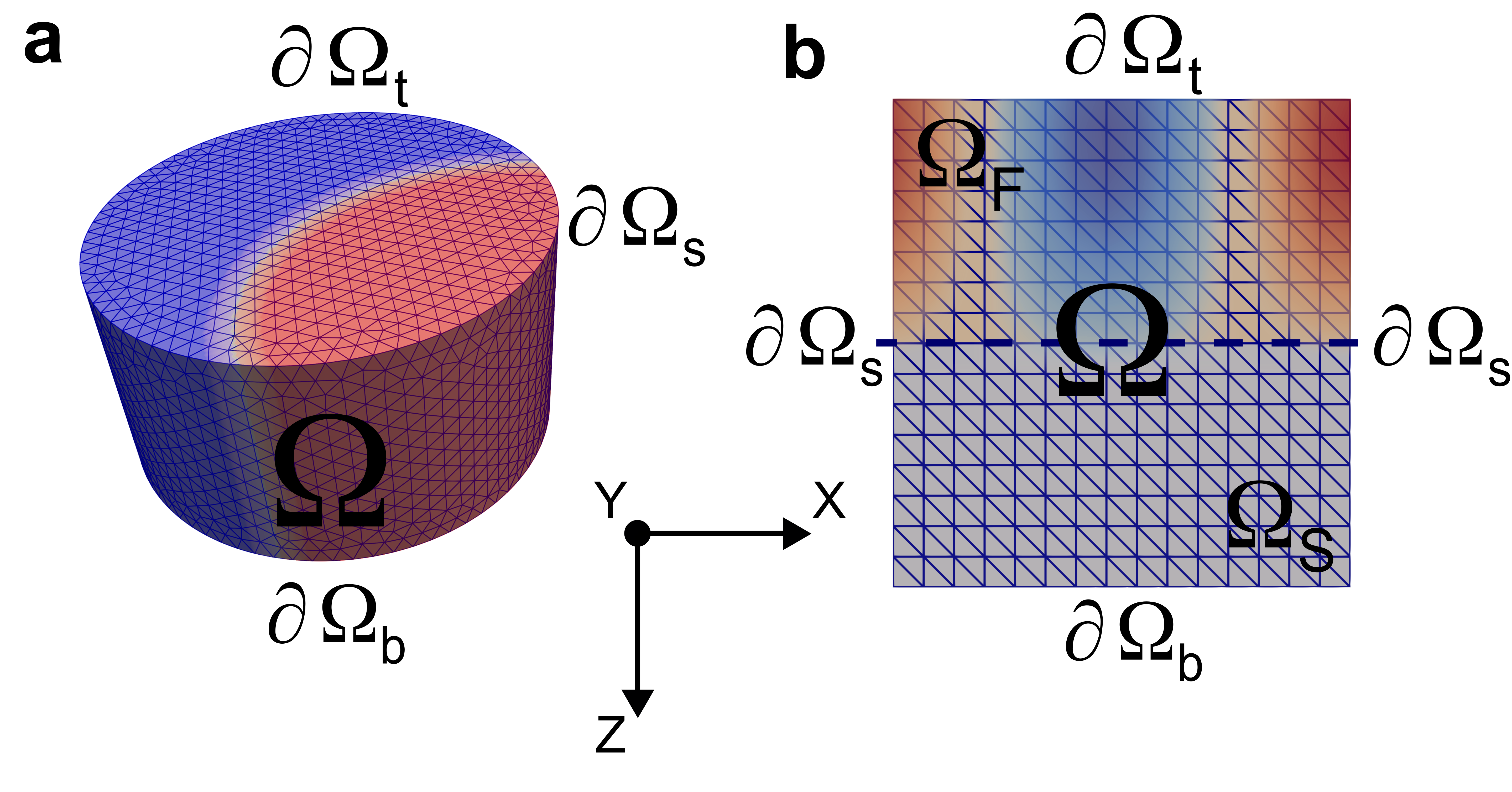}
    \caption{\small \textbf{Finite-element meshes.} \textbf{a,} Cylindrical mesh used for two-domain simulations in the MFM setup. \textbf{b,} Front-side view of rectangular mesh used for multi-domain simulations in the MFSM setup. The red and blue colors correspond to the up- and down- polarization directions, respectively.}
    \label{FigMeshes}
\end{figure}

~~\newline
\small{
The minimum of the energy functional\,(\ref{FEFunctional}) is found by solving relaxation equation
\begin{equation}
\label{eqn:func}
    -\gamma \frac{\partial \mathbf{P}}{\partial t} = \frac{\delta F}{\delta \mathbf{P}}\,,
\end{equation}
where $\delta F / \delta \mathbf{P}$ is the variational derivative of \,(\ref{FEFunctional}); the time-scale parameter $\gamma$, which does not influence the sought energy minimum is taken equal to unity.
The electrostatic Poisson equation $\varepsilon_{0} \varepsilon_{i} \nabla^{2} \varphi = \nabla \cdot \mathbf{P}$, describing the spatial distribution of the polarization, is solved on the each respective relaxation step.
 
For practical implementation of simulations, we have used the open-source FEniCS computing platform\,\cite{LoggMardalEtAl2012a}. 
To create the tetrahedral finite-element meshes we used an open-source 3D mesh generator gmsh\,\cite{Geuzaine2009}. For the case of the MFM capacitor, the computational region is a cylindrical volume $\Omega$, restricted by the side boundary, $\partial \Omega_{s}$, and by the top, $\partial \Omega_{t}$, and bottom, $\partial \Omega_{b}$, boundaries, see Fig.\,\ref{FigMeshes}a. For the case of the MFSM heterostructure, the computational region is a rectangular box $\Omega$, that includes the ferroelectric layer, $\Omega_{F}$, and the semiconducting layer, $\Omega_{S}$, see Fig.\,\ref{FigMeshes}b.
The computational region, $\Omega$, is restricted by the left, right, front and back-side boundaries $\partial \Omega_{s}$ and by the top, $\partial \Omega_{t}$, and bottom, $\partial \Omega_{b}$, boundaries,

The solutions for the polarization, $\mathbf{P}(r)$,  and electrical potential, $\varphi(r)$, distribution were sought in the functional space of the piece-wise linear polynomials. For simulation of the MFM-capacitor, controlled by charge $Q$, we use free boundary conditions for $\mathbf{P}(r)$ on the whole surface of the cylinder. At the same time, it was assumed that the electrodes produce almost uniform $z$-directed electric field, $E_{\mathit{z}}=-\partial_{\mathit{z}}\varphi$, spreading through the capacitor. The boundary constraint   
$-Q/S_f= \bar P_{\mathit{z}}+\varepsilon_{0}\varepsilon_{i}E_{\mathit{z}}$ was used at the electrode interfaces to fix the applied charge $Q$ that tunes the displacement of the DW in the spontaneously emerging two-domain structure. The bar denotes averaging over the interface surface.

For simulation of the MFSM heterostructure, the relaxation equation (\ref{eqn:func}) was solved for the ferroelectric part of the sample while the electrostatic Poisson equation was solved for the whole domain. Boundary conditions for all the variables were taken to be periodic in the $x$  direction. The size of the simulation rectangular box in $x$-direction,  corresponding to the period of the 
spontaneously emerging domain structure was considered as an energy-minimizing parameter which was optimized for each series of calculations. Boundary conditions for $\mathbf{P}$ on $\partial \Omega_{t}$ and $\partial \Omega_{b}$ 
as well as on the front- and back surface boundaries of the rectangular box were taken as free boundary conditions. The Dirichlet boundary conditions were imposed on $\varphi$ at the bottom and top surfaces of the box such that $\varphi(\partial \Omega_{b}) = -U/2$ and $\varphi(\partial \Omega_{t}) = +U/2$, to reproduce the application of the voltage $U$ to the electrodes.
The effective charge at the electrode was calculated as $Q=-S_f(\bar P+\varepsilon_0\varepsilon_{i} \bar E_{\mathit{z}})$.

To approximate the time derivative in Equation\,(\ref{eqn:func}), we used the variable-time BDF2 stepper\,\cite{Janelli2006}. The initial conditions for polarization distribution were taken to be random in the range of $-10^{-6}$ - $10^{-6}$\,C\,m$^{-2}$ for the polarization magnitude at the first time-step of simulation.  The system of the non-linear equations arising from the discretization of Equation\,(\ref{eqn:func}) was solved using the Newton-based nonlinear solver with line search and  generalized minimal residual method with the restart\,\cite{petsc-web-page,petsc-user-ref}. On each time step of the  simulation in MFSM heterostructure, the linear system of equations obtained from the discretization of electrostatic Poisson equation was solved separately using a generalized minimal residual method with restart.
}

\section*{Supplementary Note 1: Ferroelectric capacitors, negative capacitance and energy}~~\\

\begin{figure*}[h!]
\centerline{
\includegraphics [width=1.0\textwidth] {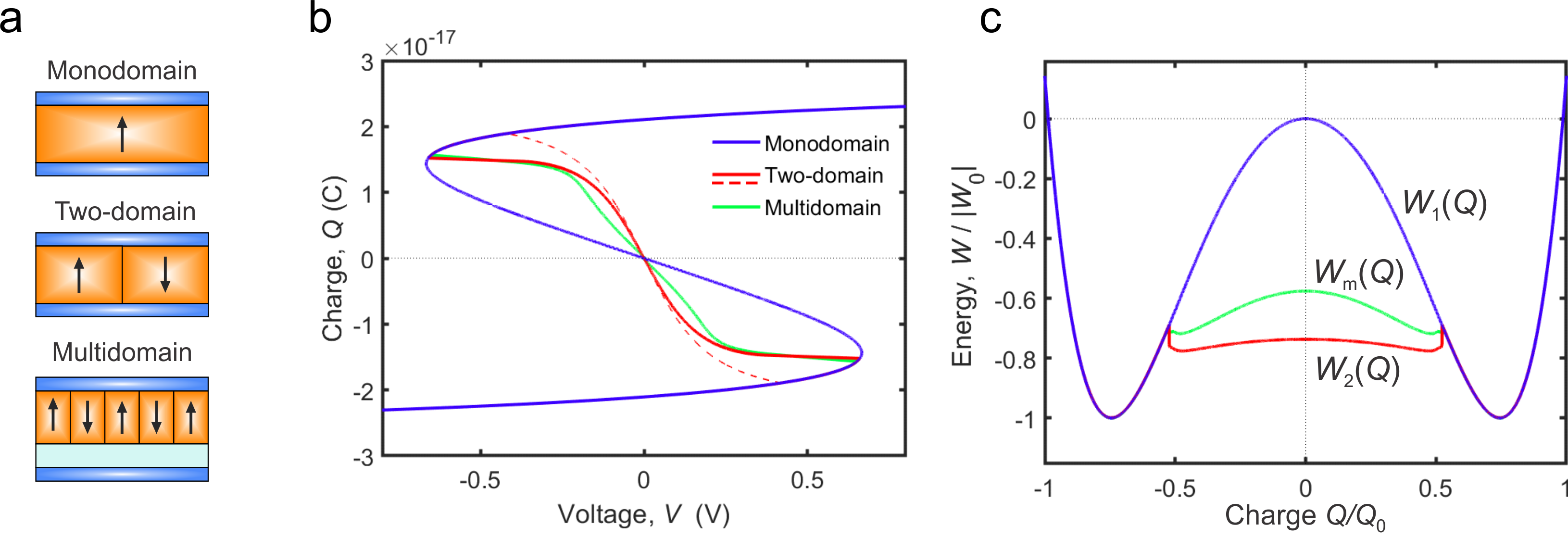}
}
\caption{\small \textbf{The characteristics of the different types of the ferroelectric capacitors}. 
\textbf{a,}\,The sketches of the domain configurations in a ferroelectric layer of  MFM and MFSM capacitors. 
From top to bottom: monodomain, two-domain  and multidomain  states, respectively. 
\textbf{b,}\,\,The charge-voltage characteristics, $Q$-$V$, of the monodomain and two-domain states of the MFM capacitors (the so-called S-curves)  
are shown by blue and red solid lines, respectively. The red dashed line corresponds to the analytical approximation for the $Q$-$V$ dependence in the two-domain state, given by Eq.\,(6) of the main text. The lime line shows the $Q$-$V$, characteristic of the multidomain state in a ferroelectric layer for the MFIM capacitor.
\textbf{c,} The double-well energy profile of the ferroelectric capacitors as function of the applied charge. The colour scheme corresponds to the states in the panel\,({b}). In case of the MFSM capacitor, only the energy fraction corresponding to the ferroelectric layer is shown. It is clearly seen that the monodomain state with energy $W_1(Q)$ is unstable against the formation of either two-domain state in the MFM capacitor with $W_2(Q)$ or the multidomain state in MFSM capacitor with $W_m(Q)$ at the fixed charge $Q$. The energy and charge are normalized on the absolute value of the energy and equilibrium charge of the monodomain short-circuited ferroelectric capacitor, $|W_0|$ and $Q_0$ respectively.
}
\label{FigCurves}
\end{figure*}

In this Supplementary Note we present the analysis of the onset of the NC in the
charge-controlled monodomain ferroelectric capacitor and demonstrate the
instability of a monodomain state against the multi-domain formation in the
capacitors with the metal-ferroelectric-metal (MFM) and
metal-ferroelectric-semiconductor-metal (MFSM) layer configurations. 
The investigated setups are sketched in \FG{Supplementary Figure\,\ref{FigCurves}a}.  \FG{Supplementary Figure\,\ref{FigCurves}b} demonstrates the
charge-voltage, $Q$-$V$, characteristics of these capacitors and \FG{Supplementary Figure\,\ref{FigCurves}c} shows the respective energy
profiles for the monodomain ferroelectric layer, $W_1$, for the two-domain state in the MFM capacitor, $W_2$,  and for the multidomain state in the MFSM capacitor, $W_m$, as functions of the applied charge $Q$. These results presenting the details of the phase field simulations give rise to dependencies shown in Figure\,1b of the main text.

In \FG{Supplementary Table\,I} for the sake of convenience, we put together our previous analytical results
\,\cite{Lukyanchuk2005,DeGuerville2005,Lukyanchuk2009,Sene2010,Lukyanchuk2019,Lukyanchuk2018}, which are in a perfect agreement with the results obtained by the simulations. In \FG{Supplementary Table\,I}
, the capacitance, $C_{f}=\varepsilon
_{0}\varepsilon _{f}\frac{S_{f}}{d_{f}}$ and the energy, $W_{0}\simeq
-S_{f}d_{f}P_{0}^{2}/\left( 8\varepsilon _{0}\varepsilon _{f}\right) $ of
the ferroelectric capacitor in the  stable monodomain polarization state
with  $P=P_{0}$, are used as the normalization parameters. The other
parameters are defined as follows: $S_{f}$ and $d_{f}$ are the capacitor
area and thickness ( $S_{f}=\pi D^{2}/4$ in case of the disk-shape capacitor
with the diameter $D$) , $\varepsilon _{0}$ and $\varepsilon _{f}$ are the
vacuum permittivity and longitudinal (along $P$) ferroelectric dielectric
constant, $\kappa =$ $\left( \varepsilon _{f}/\varepsilon _{f\perp }\right)
^{1/2}$ is the anisotropy of the dielectric tensor (where $\varepsilon _{f\perp }$ 
is the transversal dielectric constant; in Ref.\,\cite{Lukyanchuk2018} $\varepsilon _{f}$ and $%
\varepsilon _{f\perp }$ were denoted as $\varepsilon _{\parallel }$ and $%
\varepsilon _{\perp }$ respectively); $w$ is the domain width in
the multidomain texture, $\gamma _{1}$ depends on the polynomial type of the Ginzburg-Landau energy expansion and typically varies between 2 and 3, $\gamma _{2}=\frac{4}{3\pi \nu }\simeq 4.24$ (with $%
\nu \simeq 0.1$) and  $\gamma _{m}=\frac{\pi }{4\mathrm{\mathrm{ln}}2}\frac{%
\varsigma }{\kappa }\simeq 4.55$ are the numerical factors accounting for the
specific geometry and energetics of the DWs in the two- and multidomain case,
$\xi _{0}$ is the characteristic thickness of the DW, and $\varsigma 
$ is the factor, reflecting the interface boundary conditions. In our case
of the MFIM setup $\varsigma =2\left( 1+\varepsilon _{s}/\left( \varepsilon
_{f}\varepsilon _{f\perp }\right) ^{1/2}\right) $. The numerical estimates
in the \FG{Supplementary Table\,I} are given for our model cases with $d_{f}=3$\,nm, $D=6$\,nm, $\xi _{0}\approx 1$\,nm, $w\approx3$\,nm. The dielectric parameters,  
$\varepsilon _{f}=50$, $\varepsilon _{s}=25$, $\kappa =1$, $\varsigma \approx 4$, are selected to be close to those used for the conventional ferroelectric FETs with high-$\kappa$ gate dielectric and either PbTiO$_3$ or HfO$_2$-based ferroelectric materials. The capacitance $C_2$ of the two-domain MFM capacitor corresponds to $C_{NC}$ given by Eq.\,(2) of the main text. 

\medskip
\subsection{Negative capacitance of the monodomain state}~~\newline
\medskip
Here we present the main dielectric characteristics of the monodomain ferroelectric capacitor, that would have existed had this state were stable.  
 The emergence of the spontaneous polarization, $P$, in the uniaxial ferroelectric and its interaction with the
electric field, $E$, is conventionally described by the free energy functional  
\begin{equation}
\mathrm{F}=\int \left( F_{GL}(P)-EP\right) dV,  \label{Ftot}
\end{equation}
where the zero-field Ginzburg-Landau free energy is 
\begin{equation}
F_{GL}(P)=a_{3}^{*} P^{2}+a_{33}^{*} P^{4}+a_{333} P^{6}  \label{FGL}\,.
\end{equation}

For illustration purposes we take the Landau coefficients, 
$a_{3}^{*}$\,=\,$2.96\times10^8$\,C$^{-2}$m$^{2}$N$^{-1}$,
$a_{33}^{*}$\,=\,$0.05\times10^{9}$\,C$^{-4}$m$^{6}$N,
$a_{333}$\,=\,$0.26\times10^{9}$\,C$^{-6}$m$^{10}$N,
corresponding to the substrate-strained  PbTiO$_3$ at $T=25^\circ C$ with the compressive strain $-1.3\%\,$\,\cite{Pertsev1998}.

The minimization of the Eq.\,(\ref{Ftot}) with respect to $P$,%
\begin{equation}
\frac{\delta \mathrm{F}}{\delta P}=2a_{3}^{*}P+4a_{33}^{*}P^{3}+6a_{333}P^{5}-E=0,  \label{EGL}
\end{equation}
implicitly defines the constitutive relation $P=P(E)$ which is a basic relation defining
the electrodynamics of ferroelectric.

Adjusting Eq.\,(\ref{Ftot}) to the ferroelectric capacitor MFM setup,
yields the polarization-voltage relation for the monodomain MFM system
\begin{equation}
V(P)=-d_{f}E=-d_{f}\left( 2a_{3}^{*}P+4a_{33}^{*}P^{3}+6a_{333}P^{5}\right).
\label{VP}
\end{equation}%
In case of the capacitor, controlled by the charge $Q$, Eq.\,(\ref{VP}) should be
completed by the electrostatic boundary conditions at the capacitor
electrodes which are electrically isolated from the external source: 
\begin{equation}
-Q(P)=S_{f}\left( P+\varepsilon _{0}\varepsilon_i E(P)\right) =S_{f}\left(
P-\varepsilon _{0}\varepsilon _{i}\frac{V(P)}{d_{f}}\right)\,.   \label{BC}
\end{equation}
Here $S_{f}=\pi D_{f}^{2}/4$ and $d_{f}$ are the  area and thickness of the
disc-shape capacitor of diameter $D_{f}$, $\varepsilon _{0}$ is the vacuum
permittivity and $\varepsilon _{i}\simeq 10$ is the background dielectric
constant of the ferroelectric provided by the displacement of the non-polar
ions. In coherence with the cylinder geometry of NC FET considered in the article, we take $%
D_{f}\approx 6$nm and $d_{f}\approx 3$nm.

Equations (\ref{VP}) and (\ref{BC}) parametrically define the presented in
\FG{Supplementary Figure\,\ref{FigCurves}b} nonlinear $Q$-$V$ characteristics of the monodomain transistor (where 
$P$ is the running parameter) that is similar to the S-shape of the $P(E)$
constitutive relation of bulk uniform ferroelectric. 
Accordingly, the energy of the charge-driven monodomain capacitor is 
\begin{equation}
W_1=\int VdQ=S_{f}d_{f}F_{GL}(P)+\frac{1}{2}\varepsilon _{0}\varepsilon _{i}%
\frac{S_{f}}{d_{f}}V^{2}(P)\,.
 \label{W1}
\end{equation}%
The corresponding energy of the ferroelectric layer is displayed in \FG{Supplementary Figure\,\ref{FigCurves}c} as function of the capacitor charge, $Q$; using the $Q=Q(P)$
relation (\ref{BC}) it acquires a double-well profile, identical to that of the Landau
double-minima potential (\ref{FGL}). The maximum of $W_1(Q_1)$ corresponds to the
paraelectric phase with $V,Q=0$ and the two degenerate minima describe the states with $V=0$ and $\pm
Q_{0}=\pm S_{f}P_{0}$ where  $\pm P_{0}$ are the stable polarization states
of a uniform short-circuited ferroelectric slab given by the minima of Eq.\,(\ref{FGL}). 

\begin{figure*}[t!]
\centerline{
\includegraphics [width=1.0\textwidth] {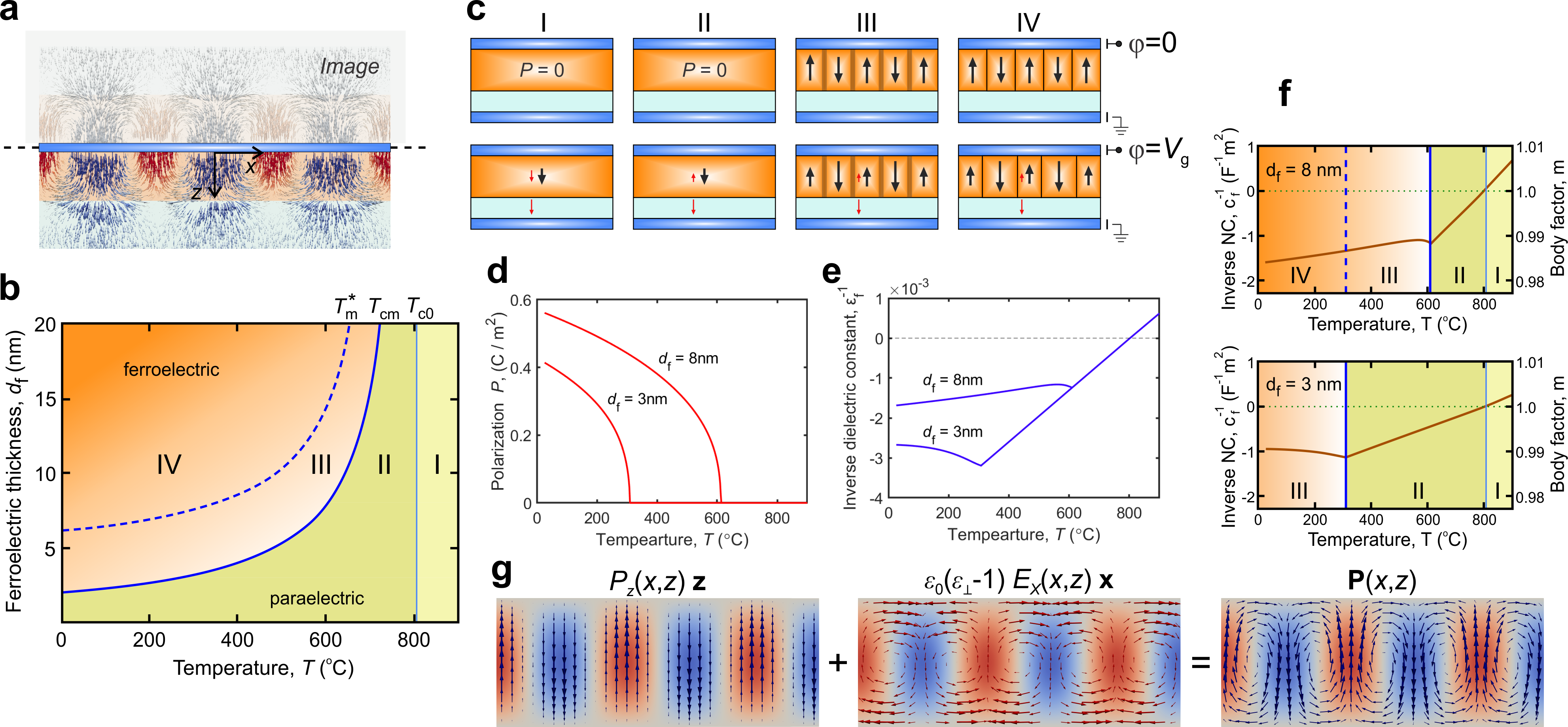}
}
\caption{\small  \textbf{Multidomain state in the MFS FET.}
\textbf{a,}\,The lower part of the panel, below the dashed line, exhibits the polarization distribution in the up (red arrows) and down (blue arrows) oriented domains in the ferroelectric (orange) and semiconducting (pale blue) layers of the MFS FET, located below the metallic (sky blue) electrode. 
The upper part of the panel displays an image of the fictitious graphic continuation of the system, visualising its reflection in the metallic electrode. It is seen that the whole, the system and its image, is effectively equivalent to the IFI heterostructure. 
\textbf{b,}\,The $d_f$-$T$ phase diagram of the MFS FET comprising four states having distinct dielectric properties: I denotes the paraelectric state, II is the incipient ferroelectric state, III is the soft-domain ferroelectric state, and IV is the hard domain ferroelectric state.
\textbf{c,}\,The response of the ferroelectric states I-IV to the applied gate voltage, $V_g$. The states I and II maintain the paraelectric phase with $P=0$ at zero voltage. Applied finite voltage induces the nonzero uniform polarization, shown by black arrow, and the internal electric fields in the ferroelectric and semiconducting layers shown by red arrows. Importantly, the direction of the electric filed in the ferroelectric layer in the incipient ferroelectric phase II is opposite to the applied voltage, which is manifestation of the negative capacitance (NC) effect.  
The inverse direction of the electric field in ferroelectric layer, hence the NC effect, maintain in the soft-domain (III) and in hard-domain (IV) phases (only the average electric filed is shown). 
The alternative change of the domains polarization amplitude (phase III and IV) and motion of the domain walls (phase IV) cause the NC dielectric response. 
\textbf{d,}\,The temperature dependence of the average polarization amplitude, $P$, for the multidomain ferroelectric layers of thickness $d_f=3$\,nm and $d_f=8$\,nm in MFS FET.
\textbf{e,}\,The temperature dependence of the inverse dielectric constant, $\varepsilon_f^{-1}$, for the same multidomain ferroelectric layers of different thicknesses.
Importantly, $\varepsilon_f^{-1}$ is negative at temperatures below the bulk critical temperature $T_{c0}$.
\textbf{f,}\,The temperature dependence of the inverse of the area-normalized capacitance, $c_f^{-1}$, for the same multidomain ferroelectric layers of different thicknesses 
and of the corresponding body factors of the respective MFS FETs with the capacitance of the semiconducting layer $c_f=0.01$\,F\,m$\red{^{-2}}$. The colored regions correspond to different ferroelectric states, shown in panel\,({c}).
\textbf{g,}\,
In the soft multidomain state in ferroelectric slab the gradually distributed spontaneous $z$-directed polarization, $P_{\mathit z}(x,z)\,\mathbf{z}$ (left) produces the electric depolarization field, which is mostly oriented in the plane of the ferroelectric layer. This field triggers the induced polarization $\varepsilon_0(\varepsilon_\perp-1) E_{\mathit x}(x,z)\,\mathbf{x}$ (middle). The superposition of the induced polarization and spontaneous polarization yields the vortex-like polarization texture $\mathbf{P}(x,z)$ (right).  $P_{\mathit z}(x,z)\mathbf{z}+\varepsilon_0(\varepsilon_\perp-1) E_{\mathit x}(x,z)\,\mathbf{x}=\mathbf{P}(x,z)$.
}
\label{FigMultiPar}
\end{figure*}

The existence of the NC in a mono-domain ferroelectric capacitor, had such a state been stable, follows from the double-well landscape of the capacitor energy $W$ as a function of the applied charge $Q$\,\cite{Landauer1976}. 
The upward curvature of $W_1(Q)$ at small $Q$ 
corresponding to the incipient ferroelectric state  with $P=0$ at $Q=0$ implies that the addition of a small charge to the ferroelectric capacitor plate, induces non-zero polarization and reduces its energy. Hence a negative value of the capacitance  $C_{1}^{-1}=(d^2W_1/dQ^2)_{Q=0}$. 
The realization of the NC effect rests on the design ensuring that it is the charge that controls the state of the capacitor\,\cite{Iniguez2019}. 
Had one allowed for the charge to vary freely, for example, by short-circuiting the capacitor's plates, the system would have fallen into one of the energy minima corresponding to the spontaneously polarized states. 
The in-series capacitance circuit of the NC FET, shown in \FG{Figure\,1a} of the main text, in which $C_g$ is replaced by $C_{NC}<0$, sets in the required charge-controlled operational mode.  
The controlling charge is set by the gate voltage, $V_g$, as $Q={C}_{FET}V_{g}$, where $C_{FET}^{-1}=C_{s}^{-1}-|{C}_{NC}^{-1}|$ is the total capacitance of the transistor.
This mode is  reversible as long as $C_{FET}$ remains positive, ensuring condition $m>0$.

\bigskip
\subsection{Instability of a monodomain state against the multi-domain formation}~~\newline
The above consideration pertains to a single-domain configuration which is widely used to explain the NC of the ferroelectric layer integrated into the FET. 
However, as follows from the fundamental thermodynamics, the homogeneous phase with $P=0$, corresponding to the maximum of the energy landscape at $Q=0$, is unstable with respect to the spinodal decomposition into the mixture of phases with oppositely oriented polarization. This mixture possessing the lower energy is a texture of the oppositely oriented ferroelectric domains.  Importantly, the integral phase-controlling parameter, which is the charge $Q$ at the electrodes in our case, conserves upon the decomposition. The instability of a monodomain state against the multi-domain formation is clearly seen from \FG{Supplementary Figure\,\ref{FigCurves}c}, in which the monodomain energy branch, $W_1(Q)$ lies above the two- and multidomain branches $W_2(Q)$ and $W_m(Q)$ for the two-domain MFMIS, and multi-domain MFS and MFIS realization of FETs. This instability naturally appears also in the course of our phase-field simulations, where the initial uniformly polarized state spontaneously relaxes towards the domain structure. The structural details of the formed coarse-grained polarization  texture, depend on the interaction with the external environment.
In the case of MFMIS FET architecture, the system falls into the two–domain state as described in the main text. In the MFS and MFIS FETs the multidomain state is formed and its structure is described in the subsequent Note.

\section*{Supplementary Note 2: Pitfalls of MFS and MFIS FETs related to the emergence of domains}~~\\

\begin{figure*}[t!]
\centerline{
\includegraphics [width=1.0\textwidth] {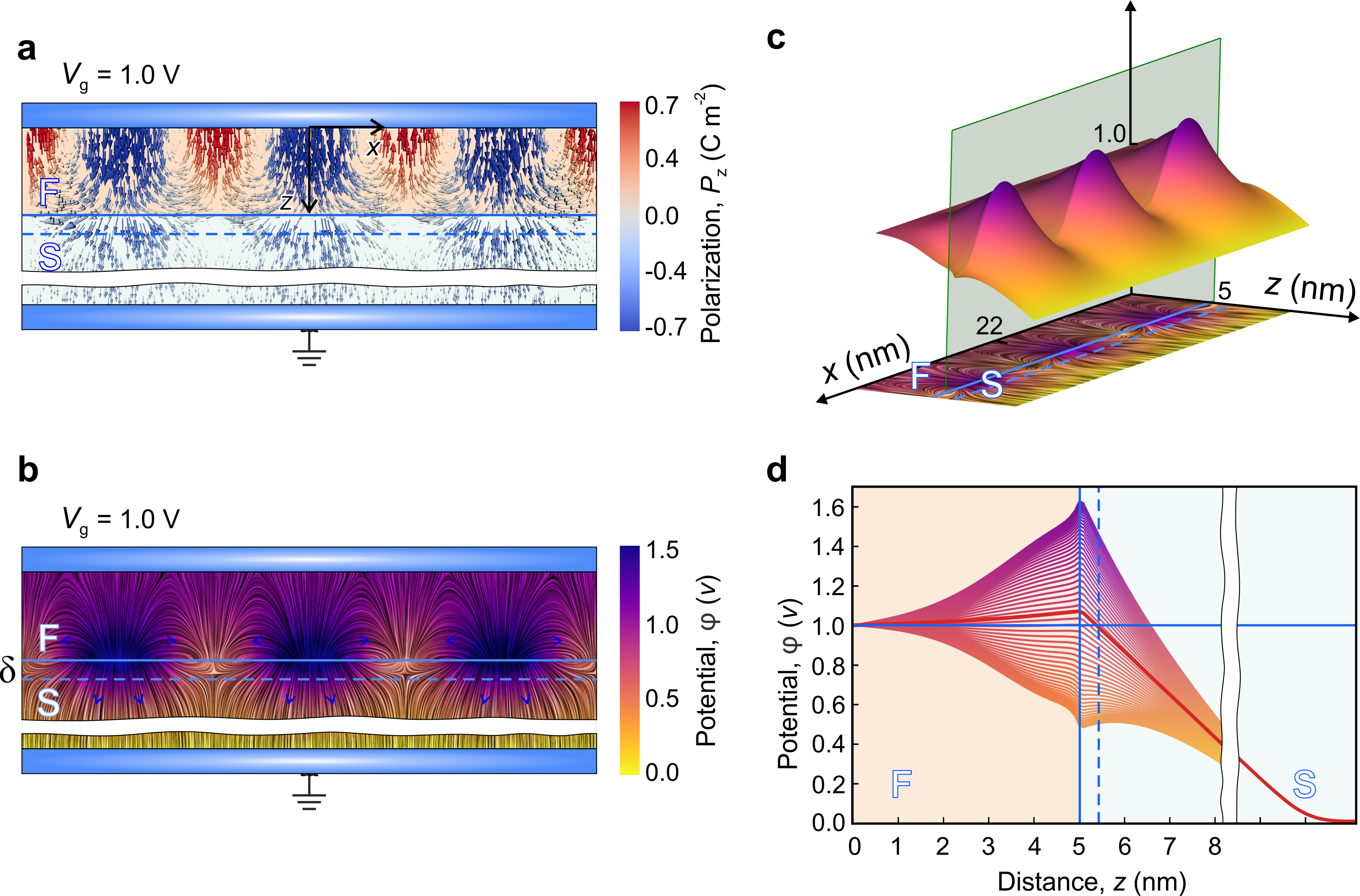}
}
\caption{\small  \textbf{The fields distribution in the multidomain state of the MFS FET under the applied gate bias}.  The potential $\varphi_0=1.0$\,V is applied to the top gate electrode. The bottom electrode is grounded. The region of the conducting channel is marked by the dashed line.
\textbf{a,}\, The distribution of polarization in domains in the ferroelectric (F) layer is highly nonuniform. This non-uniformity extends even over the semiconducting (S) layer, disturbing the near-channel region. 
An applied electric field induces the non-zero average polarization by favoring the domains oriented along the field. However, the domain polarization dispersion still exceeds the average polarization value. 
\textbf{b,}\,The electric potential and electrostatic field lines in the ferroelectric and semiconducting channel regions are highly nonuniform, similarly to the polarization distribution.  
\textbf{c,} The 3D plot of the nonuniform electric potential distribution induced by the ferroelectric domains at the FS interface.
\textbf{d,}\,\,The average potential distribution (red line)  across the capacitor and the domain-induced potential scatter. The average potential enhancement at the FS interface and at the near-channel region with respect to the applied $\varphi_0=1.0$\,V corresponds to the NC effect. However, the scattering in the potential in this region substantially exceeds the potential enhancement that perturbs the conductivity channel, making the MFS configuration of the NC FET impractical. 
(The images are produced with Surface LIC toolbox in data visualisation application Paraview\,\protect\cite{ASPC2015})
}
\label{FigMultidom}
\end{figure*}

Although the NC effect has already been experimentally demonstrated in the
dielectric-insulator superlattices\,\cite{Zubko2016} and although the MFS and MFIS gate stacks  
are being considered as most promising candidates for the NC applications\,\cite{Hoffmann2021}, we show here that these structures can hardly be utilized for building the NC FET in practice. The major difficulty is the
already mentioned destabilizing action of the depolarization fields,
produced by the depolarization charges arising at the FI and FS interfaces
that results in compulsory formation of the nonuniform domain texture.\ 
Although the domain structure keeps the collective NC effect\,\cite{Lukyanchuk2018}, \textbf{two important
offshoots inhibit its  practical implementation in the MFS and MFIS FET}. Those are:
\begin{itemize}
  \item The nonuniform field distribution 
disturbs the conducting channel formation in vicinity of FS interface.
  \item The high absolute value of the NC  cannot properly match with the
semiconducting substrate. 
\end{itemize}
 
To bury the structures having these pitfalls on the
quantitative basis, we note that the electrode-ferroelectric interface is
described by the electrostatic equipotential boundary condition, hence the MFS
system is equivalent to the dielectric-ferroelectric-dielectric (IFI)
heterostructure in which the thickness of the ferroelectric layer is
doubled, as shown in \FG{panel\,(a)} of \FG{Supplementary Figure\,\ref{FigMultiPar}},
and the electrostatic properties of dielectric layers are described by the
dielectric constant of the depleted region of semiconductor, $\varepsilon
_{s}$. This mapping allows us to use the results and calculation techniques developed in\,\cite{Lukyanchuk2005,DeGuerville2005,Lukyanchuk2009} for the IFI 
heterostructures and compare them with results of the phase-field simulations described in 
Methods Section of the main text.

Deriving from the results of\,\cite{Lukyanchuk2005,DeGuerville2005,Lukyanchuk2009} we note, first, that the transition temperature to the
multidomain ferroelectric state of MFSM structure, $T_{cm}$, is lower then that for the monodomain
bulk sample with the short-circuited electrodes, $T_{c0}$, and is given by
\begin{equation}
T_{cm}=\left( 1-2\pi \left( \frac{C_{Curie}}{T_{c0}\varepsilon _{\perp }}%
\right) ^{1/2}\frac{\xi _{0}}{2d_{f}}\right) T_{c0},  \label{Tcm}
\end{equation}%
where $C_{Curie}$ is the Curie constant in the paraelectric phase and factor $2$
before $d_{f}$ corresponds to the mentioned effective thickness doubling. The inverse
scaling of the renormalized  transition temperature with ferroelectric layer
thickness, $d_{f}$, allows us to sketch the temperature-thickness phase
diagram of the system, presented in \FG{panel\,(b)} were we distinguish
four typical regions for states with distinct dielectric and ferroelectric
properties, depicted in \FG{panel\,(c)}. The dependencies of the average polarizations, dielectric
constants and capacitancies for the ferroelectric layers of different thickness
on the temperature  are shown in \FG{panels (d-f)} respectively.

I. Paraelectric state at $T>T_{c0}$ . The system stays at the paraelectric
state with the positive dielectric constant $\varepsilon _{p}$, hence
capacitance $C_{p}$ is given by the Curie relation 
\begin{equation}
C_{p}=\varepsilon _{0}\varepsilon _{p}\frac{S_{f}}{d_{f}},\quad \varepsilon
_{p}=\frac{C_{Curie}}{T-T_{c0}}+\varepsilon _{i}.  \label{EpsC}
\end{equation}

II. Incipient ferroelectric state\,\cite{Iniguez2019} existing at  $T_{c0}>T>T_{cm}$. This
state precedes the transition to the multidomain ferroelectric state, however,
the system maintains the paraelectric phase with $P=0$ at $V_g=0$.
Importantly, the value of the dielectric constant is still given by the relation
(\ref{EpsC}).  After passing through the pole at $T_{c0}$, the dielectric
constant becomes negative and decreasing by its absolute value from infinity at $%
T_{c0}$ to the minimal possible (for the incipient state) value at $T_{cm}$\,\cite{Sene2010,Cano2010}. 
That is why, the incipient
state is also considered as a potential state for realizing the NC
FET\,\cite{Cano2010}. The origin of the NC in the incipient state is the specific
arrangement of the depolarization charges at the FS interface, that inverts the induced
intrinsic field within the ferroelectric layer oppositely to the applied
voltage. The corresponding NC, calculated at $T_{cm}$ is:
\begin{equation}
C_{ins}\approx -\varepsilon _{0}\frac{S_{f}}{d_{f}}\frac{1}{\pi }\left(
\varepsilon _{\perp }\frac{C_{Curie}}{T_{c0}}\right) ^{1/2}\frac{\xi _{0}}{%
d_{f}}.  \label{Cins}
\end{equation}%

III. Soft multidomain ferroelectric state emerging at $T\leq T_{cm}$. The domain
structure, shown in \FG{\ panel\,(g)} has a gradual periodic polarization
profile composed of the spontaneous $z$-component $P_{\mathit z}\sim \cos \frac{%
\pi z}{2d_{f}}\cos \frac{\pi x}{w}$ and \ the depolarization-field induced
component, that in the soft-domain state has, preferably, the $x$ orientation $%
P_{\mathit x}=-\varepsilon _{0}\left( \varepsilon _{\perp }-1\right) \partial
_{\mathit x}\varphi $\thinspace $\sim \sin \frac{\pi z}{2d_{f}}\sin \frac{\pi x}{w}$. 
Here the coordinate axis $x$ and $z$ are selected along and across the
ferroelectric layer with the origin at the FM interface and the
depolarization potential is given by Poisson equation: $\varepsilon
_{0}\varepsilon _{f\perp }\partial _{\mathit x}^{2}\varphi +\varepsilon
_{0}\varepsilon _{i}\partial _{\mathit z}^{2}\varphi =\partial _{\mathit z}P_{\mathit z}$. The
overall soft polarization profile, appearing as a superposition of the spontaneous and
field-induced components, $\mathbf{P}=\left( P_{\mathit x},P_{\mathit z}\right) $, (see 
\FG{panel\,(g)}) is seen in experiment as a periodic vortex-antivortex
texture and was also dubbed as a vortex phase\,\cite{Yadav2016}. The negative capacitance in the
soft-domain phase is provided by the dominant phase volume of wide
domain-wall regions\,\cite{DeGuerville2005}  having the locally negative dielectric constant\,\cite{Iniguez2019}. Its absolute value continues to slightly decrease, see also Refs.\,\cite{Lukyanchuk2009,Sene2010,Zubko2016} for the interpolation formulas.

IV.\ Hard multidomain ferroelectric state, stabilizing below crossover temperature at $T_m^*<T_{cm}$ with
the Landau-Kittel like flat-polarization domain profile. The mechanism of NC
is mostly provided by the field-induced motion of domain walls\,\cite{Lukyanchuk2018}. Although
this regime is not fully reachable in the nanometer-thin films (see phase
diagram at \FG{panel\,(b)}), the respective value of capacitance,  $C_{m}<0$, calculated 
in\,\cite{Lukyanchuk2018} and given in 
in the \FG{Supplementary Table\,I}, can be considered as the lower bound for the absolute
value of NC of ferroelectric layer in MFS and MFIS FETs.  

\FG{Supplementary Figure\,\ref{FigMultidom}} demonstrates the distribution of the 
polarization and electric field in the MFS FET with the $d_{f}=8\,$nm thick
ferroelectric layer at room temperature, the
voltage applied at the gate being $V_{g}=1\,$V. The FS interface and the
region of induced conducting channel, spaced from the interface on 
$\delta $, are shown by the solid and dashed blue lines, respectively.  Importantly,
the nonuniform distribution of polarization in the ferroelectric, shown in 
\FG{panel\,(a)} induces the nonuniform distribution of the electric field
and potential (\FG{panels (b) and (c)}) that penetrate into the
semiconductor over the penetration depth $\sim w$. Therefore, the
distribution of the depolarization electric potential in the channel region, 
$\varphi _{s}\sim\cos \frac{\pi x}{w}e^{-\pi \delta /w}$, is
highly nonuniform. The characteristic variation of $\varphi _{s}$ in the
near-channel region $\sim 1.2$V, shown in \FG{panel\,(d)} by the
potential line scattering is remarkably big. Not only it does exceed the voltage
gain in the near-channel region $\sim 0.3$V due to ferroelectric NC,
shown by the read line,  but is also comparable with the transistor
operation voltage. This strong nonuniformity of the field makes the
conducting channel inoperable. It also implies that the suggestion to introduce
the buffer dielectric layer in between the ferroelectric and semiconducting
layers to attenuate the field oscillations, making thus the MFIS
configuration, will barely improve the situation. As it is clear from the
plots of \FG{panel\,(d)}, the damping of the field nonuniformity occurs at
the distances $>w$ where the NC potential amplification effect is not actual
anymore.

Another problem related to the realization of the MFS FET, is the very big absolute value of the surface-normalized NC of the ferroelectric gate, $c_{NC}$, which, typically, does not go below $0.6$\,F\,m$^{-2}$, see \FG{Supplementary Figure\,\ref{FigMultiPar}}, which substantially exceeds the capacitance of semiconducting substrate $c_s\sim0.01$\,F\,m$^{-2}$.  Such a relation between $c_{NC}$ and $c_s$ does not allow to achieve the visible gain in the amplification of the input voltage in the channel region since the decreasing body factor below $m=1$ is very small, as it is shown at the same panel. This problem is especially pertinent for the  attempts to realize the NC in the incipient regime, which is seemingly attractive due to the homogeneous distribution of the electric potential in the region of the conducting channel.  However, at these parameters the body factor is not going to reduce below 0.99\,\cite{Cano2010}, as can be also estimated from the formula (\ref{Cins}) which, together with the difficulty in achieving the proper temperature regime, makes the idea unpractical. 


\bibliographystyle{naturemag}
\bibliography{NCFET}

\vspace{-0.3cm} 
\section*{Acknowledgements}~~\newline
This work was supported by H2020 RISE-MELON action (I.L.), and by Terra Quantum AG (I.L., A.R., and V.M.V.). The work of V.M.V. was supported in part by Fulbright Foundation.

\end{document}